\documentclass[12pt]{article}

\usepackage{graphicx}% Include figure files
\usepackage{bm}% bold math
\usepackage{amssymb}
\usepackage{enumerate}

\usepackage{graphicx} 
\usepackage{cite}
\usepackage{wrapfig}
\usepackage{graphicx}
\usepackage{amssymb}
\usepackage{amsfonts}
\usepackage{amsmath}
\usepackage{longtable}
\usepackage{rotating}
\usepackage{lscape}
\usepackage{epsfig}
\usepackage{multirow}

\begin{document}

\title{Search for Light Dark Matter with accelerator and direct detection experiments: comparison and complementarity of recent results}
%The comparison of accelerator NA64, BaBar   and underground experiments on the search for light dark matter} 
\author{  S.~N.~Gninenko$^1$ \thanks{Sergei.Gninenko@cern.ch},
D.~V.~Kirpichnikov$^1$ \thanks{kirpich@ms2.inr.ac.ru},  
  N.~V.~Krasnikov$^{1,2}$ \thanks{ Nikolai.Krasnikov@cern.ch} 
  \\
\\ \normalsize $^1$ Institute for Nuclear Research RAS, Moscow 117312, Russia
\\
\\ \normalsize $^{2}$ JINR, Dubna, Russia
}
%\date{October,1997}
\maketitle
\begin{abstract}
  We discuss the most sensitive constraints on Light Dark Matter (LDM) 
  from accelerator experiments  NA64 and  BaBar   and compare it  with recent results from direct  searches at XENON1T, 
  DAMIC-M, SuperCDMS, and DarkSide-50.
  We show that for the dark photon ($A'$) model with scalar LDM,   NA64
  gives more stringent bounds for $A'$ masses $m_{A'} \leq 0.15~GeV$ than
  direct searches. Moreover,  for the case of
  Majorana LDM the damping DM velocity $v$  factor,  $v^2 \sim O(10^{-6})$, 
  for the elastic LDM electron(nucleon) cross section
   makes direct observation of Majorana LDM extremely challenging, while the absence of   this suppression in the 
  NA64 case gives an advantage to the experiment.
  The similar situation takes place for pseudo-Dirac LDM. The BaBar provides the most stringent bounds 
  for $A'$ masses $m_{A'} \geq 0.35~GeV$. For scalar LDM the direct detection experiments give
  more stringent bounds at  $m_{A'} \geq 0.35~GeV$ while
  for Majorana and pseudo-Dirac LDM case, the BaBar bounds are more stringent. The complementarity of the two 
  approaches in searching for LDM is underlined. 
\end{abstract} 

\newpage
\section{Introduction}
At present the most striking evidence in favour of new physics beyond the Standard Model (SM) is the
existence of Dark Matter(DM) in the Universe. The nature of DM is one of the
most important questions  in modern science. Among numerous  candidates for the
role of particle DM \cite{Rubakov:2017xzr, Kolb:1990vq},  light DM (LDM)   with the mass of
DM particles $\chi$ below the electroweak scale are of great interest.
In the most popular model, a  massive  vector boson, called dark photon $A'$,
 \cite{Okun:1982xi,Holdom:1985ag,Pospelov:2007mp} interacting  with LDM particles in a similar way as the ordinary photon interacts with charged particles in the SM.
The mixing of the $A'$  with the photon leads to a new feeble interaction of
SM  leptons and  quarks with LDM particles, allowing LDM  and visible matter to be  initially in thermal equilibrium during the early universe 
via  annihilation into each other at equal rates.  With this assumption 
 it is possible to  predict the decoupling cross section of LDM from ordinary  matter and the observed relic  density of DM in our universe.
This in turn, results in  relations among  four  essential parameters - mixing strength $\epsilon$, dark
photon mass $m_{A`}$, LDM mass $m_{\chi}$ and the interaction coupling
constant $e_D (\alpha_D \equiv \frac{e_D^2}{4\pi}) $ of the dark photon with LDM.  Note, that the $A'$ interacts with charged particles similar to the ordinary photon with coupling constant $e' = \epsilon e$. 
Surprisingly, in this framework LDM could be as light as  $m_\chi \leq ~O(1)~GeV$, and the above mentioned parameters lie in the region 
which can be probed with accelerator-based  and  direct detection experiments 
\cite{Boehm:2003hm, Essig:2013lka,Alexander:2016aln,Battaglieri:2017aum,Beacham:2019nyx,Alemany:2019vsk,Agrawal:2021dbo,Essig:2022yzw,Batell:2022dpx,Gori:2022vri,Lanfranchi:2020crw,Fabbrichesi:2020wbt,Berlin:2018bsc,Boveia:2022syt,Gninenko:2020hbd, Gninenko:2021csg,Antel:2023hkf,Dutta:2023fij}. 

The benchmark parameter
$$
\alpha_{D} \simeq 0.02 f \left( \frac{10^{-3}}{\epsilon} \right)^2 
\left( \frac{m_{A'}}{100 \, \mbox{MeV}} \right)^4 \left( \frac{10 \, \mbox{MeV}}{m_{\chi}}\right)^2 
%y= \epsilon^2 \alpha_D (m_{\chi}/m_{A'})^2
$$
is typically used for comparing the experimental sensitivities to the parameter space of the relic DM 
density~\cite{Kolb:1990vq, deNiverville:2011it,Izaguirre:2014bca, Antel:2023hkf, Alexander:2016aln,Gninenko:2020hbd,Gninenko:2021csg}. Here,  $f \simeq \mathcal{O}(1)$ is the coefficient 
that depends on  the specific type of DM.  It should be noted that dark photon model is a renormalizable 
model,   with the scalar, Majorana, and pseudo-Dirac LDM  cases often discussed. 
%Namely the relation $\epsilon^2 \alpha_D =f(m_{\chi}, m_{A'})$ 
%C \frac{(m^2_{A'}- 4 m_{\chi}^2)^2}{m^2_{A`}}$
%\cite{FORMULA,Gninenko2020, Gninenko2021}  

%is often used. Here $f(m_{\chi}, m_{A'}) $ - is known function which
%depends on the nature of LDM. 

  %Moreover, for dark photon model
  %the annihilation of LDM into obsevable matter takes place via $A'$  s-channel
  %and $m_{A'} \geq m_{\chi}~$  \cite{FORMULA, Gninenko2020, Gninenko2021}.

There are at least two approaches  for the  search for the LDM particles.
In the first one,  accelerator  experiments are used for the indirect searches for
LDM. The typical examples are the fixed-target NA64 experiment at CERN,  and BaBar experiment at $e^+ e^-$ collider at SLAC \cite{Alexander:2016aln,Gninenko:2020hbd, Gninenko:2021csg}.
In the second approach, direct searches for signals from collisions of DM particles
with nuclei or electrons of a target at  
direct detection experiments such  as XENON1T \cite{XENON:2019gfn}, 
  DAMIC-M~\cite{DAMIC-M:2023gxo}, SuperCDMS~\cite{SuperCDMS:2023sql}, and DarkSide-50~\cite{DarkSide-50:2023fcw} is used.  Therefore it would be of great interest to compare the results
of the search for LDM particles for accelerator  and direct detection experiments. 
It is clear that  the result of the comparison depends on concrete LDM model, thus making the constraints from these two type of searches highly complementary to each other.
In  this note,  we report the results of such a comparison for the $A'$ model.  Interestingly, we found that  in some cases the accelerator
bounds are stronger than the corresponding  bounds from direct searches. 

The paper is organized as follows. In Sec.~\ref{AccBoundSect} we discuss the bounds on LDM from 
accelerator-based experiments. In Sec.~\ref{DirectBoundSec} we compare corresponding limits with the 
constraints from the direct detection experiments. We conclude in Sec.~\ref{ConclSect}.  
  \begin{figure}[t]
\begin{center}
\includegraphics[width=0.6\textwidth]{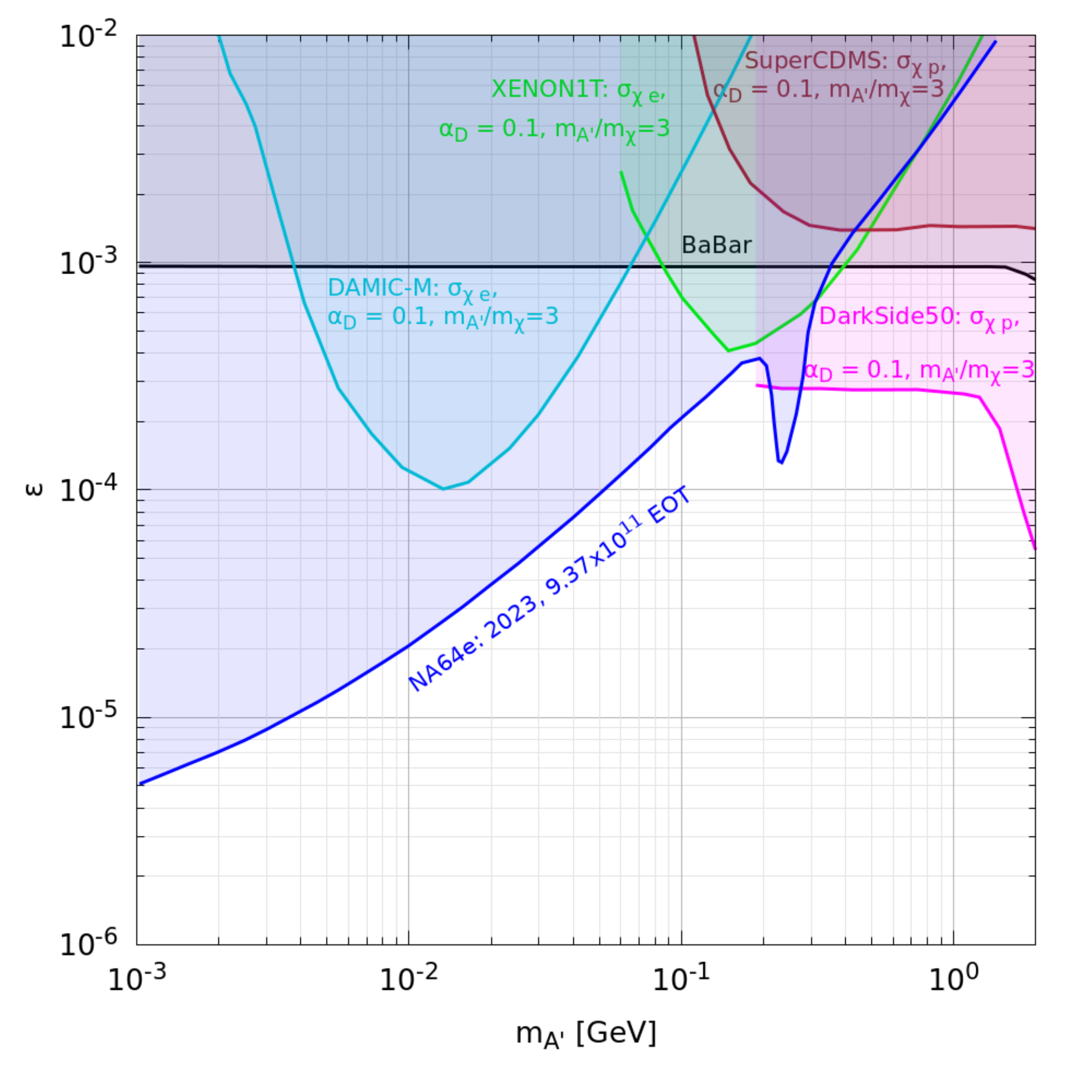}
\vspace{-3mm}
\caption{The bounds in the $(m_{A'};\epsilon)$ plane  from NA64~\cite{NA64:2023wbi} and  BaBar~\cite{BaBar:2017tiz}.  Constraints from  direct search experiments  XENON1T~\cite{XENON:2019gfn}, DAMIC-M~\cite{DAMIC-M:2023gxo}, SuperCDMS~\cite{SuperCDMS:2023sql},  and DarkSide50~\cite{DarkSide-50:2023fcw} derived for scalar/Dirac dark matter  and 
 benchmark  point  $\alpha_D = 0.1$, $\frac{m_{A'}}{m_{\chi}} = 3$ are also shown. }
\end{center}
\label{fig01}
%\labelf{fig01}
\vspace{-5mm}
\end{figure}

\section{Accelerator bounds
\label{AccBoundSect}}
At present the most stringent accelerator bounds  
were obtained from  the NA64 and BaBar experiments that searched for the LDM production in invisible decay mode of the $A'$ mediator, $ A' \rightarrow \chi \bar{\chi}$, 
assuming $2 m_\chi <  m_{A'}$.
\par For the $A'$ production,  the NA64 experiment at CERN    \cite{NA64:2016oww,NA64:2017vtt}
uses the bremsstrahlung reaction of the  high-energy electron scattering off nuclei \cite{Bjorken:2009mm} 
\begin{equation}
  e Z \rightarrow e Z A'  \,
\end{equation}
and the resonant  annihilation of high-energy positrons with atomic electrons \cite{Marsicano:2018glj,Andreev:2021fzd}
\begin{equation}
  e^+ e^- \rightarrow A'  \,
\end{equation}
followed by the prompt invisible decay of the $A'$.
New NA64  constraints on the $\gamma - A'$ mixing strength $\epsilon$ were obtained recently with the accumulated record $\approx 10^{12}$ electrons on target (EOT)~\cite{NA64:2023wbi}. 
%In the dark photon model dark photon interacts with electron as ordinary photon but with coupling constant $e' = \epsilon e$. 
The NA64 90\% C.L. exclusion region in the ($m_{A'}, \epsilon$) 
plane~\cite{NA64:2023wbi} is  shown in Fig. 1. Constraints obtained  by BaBar ~\cite{BaBar:2017tiz}  using the reaction $e^+e^- \rightarrow \gamma~A'$, $A' \rightarrow \chi  \bar{\chi} $ as a source of LDM, and from direct search  experiments are also shown. 
As one can see,   the  NA64 constraints are the most sensitive for  $m_{\chi} \leq 0.3$~GeV. Note, that NA64 and BaBar limits do not depend on the type of LDM. 

  \begin{figure}[t]
\begin{center}
\includegraphics[width=0.6\textwidth]{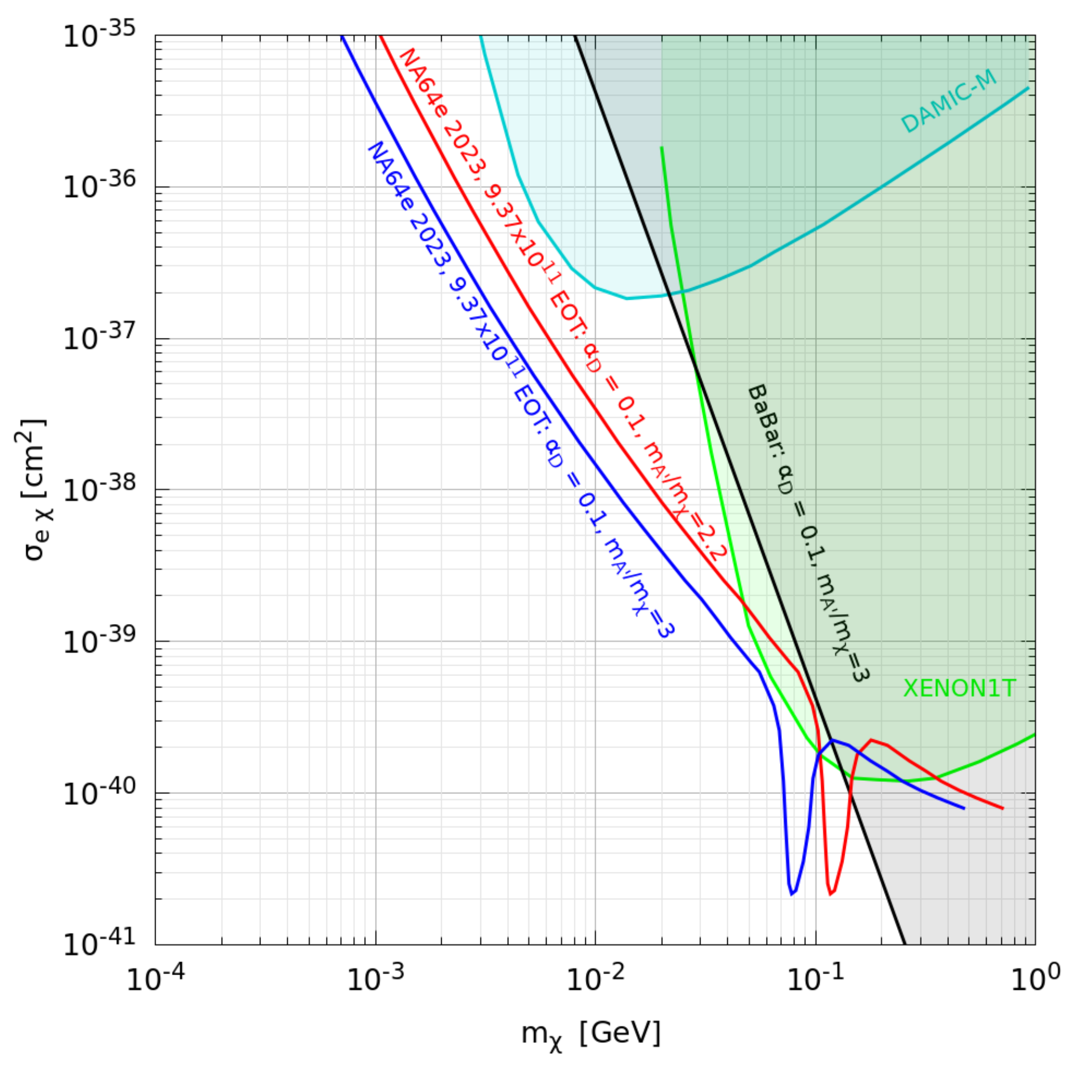}
\vspace{-3mm}
\caption{ Bounds 
  for the $A'$ model in the $(\sigma_{e\chi}, ~m_{\chi})$ plane obtained from the results of  
  NA64~\cite{NA64:2023wbi} and  BaBar ~\cite{BaBar:2017tiz} experiments. Constraints from the direct detection
  experiments   XENON1T~\cite{XENON:2019gfn} and  DAMIC-M~\cite{DAMIC-M:2023gxo}  searching for LDM in the $\chi-e$ scattering derived for the scalar and Dirac DM case and 
   benchmark points  $\alpha_D = 0.1$, $\frac{m_{A'}}{m_{\chi}} = 3(2.2)$ are also shown 
 for comparison. One can see, the NA64 bounds for masses  $m_{\chi} \leq 0.04~GeV$ are most stringent.  }
\end{center}
\label{fig01}
%\labelf{fig01}
\vspace{-5mm}
\end{figure}

\section{Direct LDM detection experiments
\label{DirectBoundSec}}
Direct detection experiments like XENON1T~\cite{XENON:2019gfn} and DAMIC-M~\cite{DAMIC-M:2023gxo} searched  for  LDM in
the elastic  $\chi$  - electron scattering,  $  \chi e \rightarrow    \chi e $, 
and also in the  $ \chi ~nucleon \rightarrow \chi ~nucleon $
scattering.  In the non-relativistic approximation for the $A'$ model, the
 spin independent  cross section $\sigma(\chi e \rightarrow \chi e)$ for the LDM is determined by \cite{Alexander:2016aln}
\begin{equation}
  \sigma(\chi e \rightarrow \chi e) = \frac{16\pi \alpha \epsilon^2 \alpha_D \mu^2_{\chi e}}{m^4_{A'}} \,,
  \label{3a}
\end{equation}
where $\mu_{\chi e} = \frac{m_{\chi} m_e}{m_{\chi} + m_{e}}$  and  $\alpha \equiv \frac{e^2}{4\pi} = \frac{1}{137}$. $\footnote{The interaction of dark photon $A'$ with Dirac dark matter $\chi$  has the form $L_{A'\chi} = e_D A'_{\mu}\bar{\chi}\gamma^{\mu}\chi $ }$. It is worth noticing that in the non-relativistic regime 
the typical cross section is given by Eq.~(\ref{3a}) for both Dirac and scalar LDM.
For Majorana LDM particle $\chi_M$ the elastic cross section
 given by 
\begin{equation}
\sigma(\chi_M e \rightarrow \chi_M e) = \frac{16\pi \alpha \epsilon^2 \alpha_D \mu^2_{\chi e}}{m^4_{A'}} k_M \,, 
\label{3b}
\end{equation}
is suppressed by the factor $k_M = \frac{2\mu_{\chi e}^2}{m^2_{\chi}}v^2_{\chi}$ \cite{Alexander:2016aln}.
Here $v_{\chi} \approx 10^{-3}$ is the velocity of LDM particle
in natural units\footnote{In natural units the speed of light $c \equiv 1 $.}.
For LDM proton elastic scattering, the corresponding formulae coincide
with (\ref{3a},\ref{3b}) up to the proton  formfactor  $F(q^2)$,  which with good accuracy is $\simeq 1$ for the LDM particles  due to a small 
transfered momentum $q \simeq m_{\chi} v_{\chi} \leq O(1)~MeV$. The results of the
comparison of NA64~\cite{NA64:2023wbi} and  BaBar~\cite{BaBar:2017tiz} constraints
with those from XENON1T~\cite{XENON:2019gfn}, DAMIC-M~\cite{DAMIC-M:2023gxo}, SuperCDMS~\cite{SuperCDMS:2023sql},
 and DarkSide-50\cite{DarkSide-50:2023fcw},   for scalar and Dirac LDM particles are presented in Figs.1-3.  For $m_{A'} > 2 m_{\chi}$ the  accelerator bounds
 are stronger than direct ones  for masses $m_{\chi} \leq  200 ~MeV$. 
 Note,  that for  the case LDM to be  Majorana particles, a  damping factor $k_M = \frac{2m_e^2}{m_{\chi}^2} \cdot O(10^{-6})$ arises, and  
  as a result,  the bounds from direct searches are very weak and can't compete with accelerator ones.

  Analogous constraints are valid for LDM models with the pseudoscalar (P) or scalar (S) mediator of a new interaction between the dark and visible sectors. 
  The NA64 experiment in this case severely limits the existence of this type of new interactions with an  electron  \cite{NA64:2021xzo}.
   Namely, for the interaction Lagrangians
  \begin{equation}
    L_S = g_S \bar{\psi}\psi S \,,
  \end{equation}
\begin{equation}
  L_P = ig_P \bar{\psi}\gamma_5\psi S \,
\end{equation}
the NA64 bounds for the coupling constant $\epsilon_S = \frac{g_S}{e}$ and  $\epsilon_P = \frac{g_P}{e}$ with electron 
%\footnote{Here $e$ is an electron charge,
%$\frac{e^2}{4\pi} \equiv \alpha = \frac{1}{137}$}
  are weaker by the factor $\sqrt{2}$ than the corresponding  bound on $\epsilon$
  for the $A'$ case for masses  $m_S, m_P \gg m_e$ (see, e.~g.~Ref.~\cite{Liu:2017htz,Liu:2016mqv} and references therein). Therefore,  the NA64 bounds for 
  the $S$ or $P$  mediators are also more restrictive
  than  the bounds for the mediator masses  $\lesssim O(100)~MeV$.
   \begin{figure}[t]
\begin{center}
\includegraphics[width=0.6\textwidth]{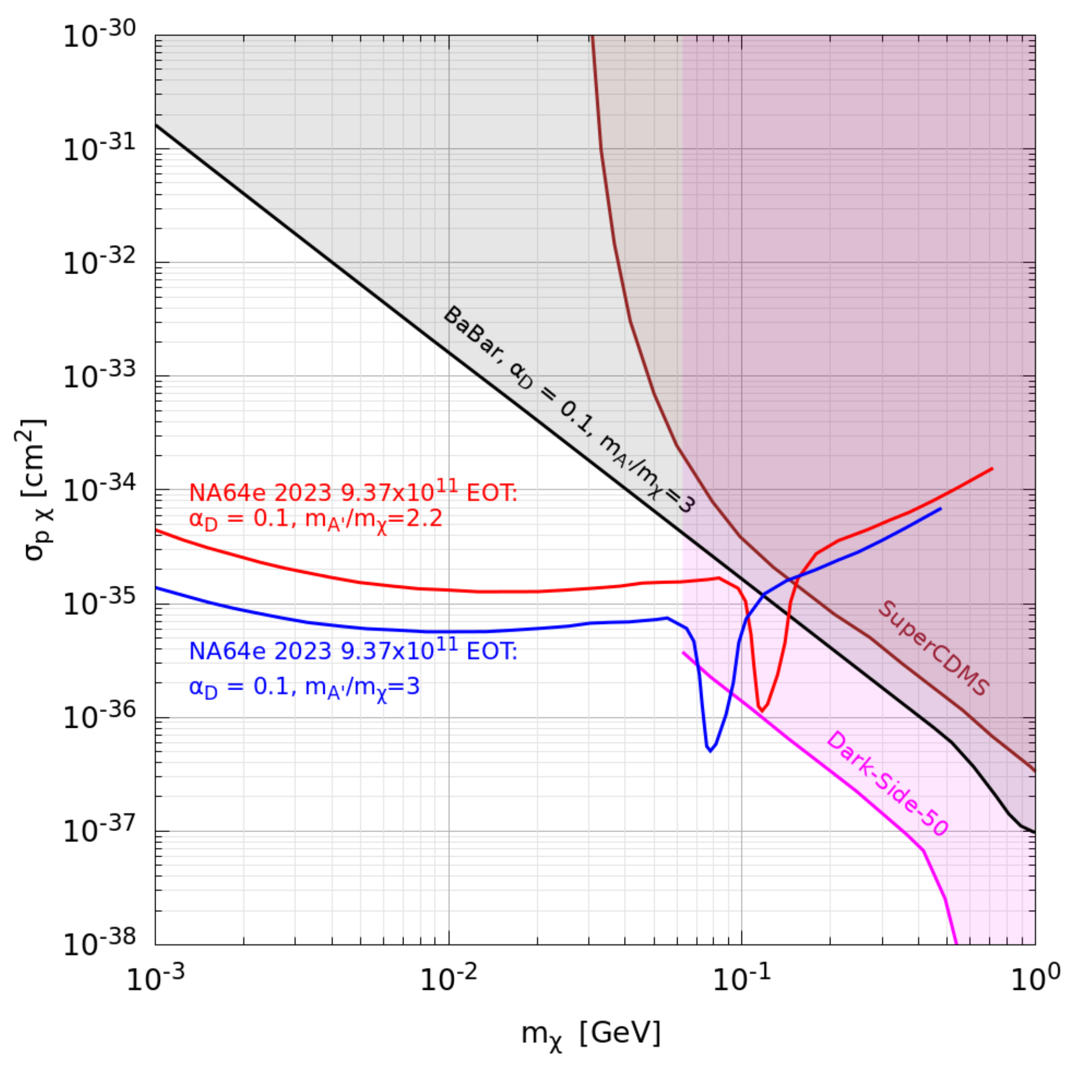}
\vspace{-3mm}
\caption{Comparison of bounds
  for the $A'$ model in the $(\sigma_{p\chi}, ~m_{\chi})$ plane for the  NA64~\cite{NA64:2023wbi},  
  BaBar~\cite{BaBar:2017tiz}, SuperCDMS~\cite{SuperCDMS:2023sql},  and DarkSide50~\cite{DarkSide-50:2023fcw} experiments. 
   The SuperCDMS and DarkSide50 limits are derived from the searching for LDM in the $\chi- p$ scattering,   
   for the scalar and Dirac DM case and  reference points $\alpha_D = 0.1$, $\frac{m_{A'}}{m_{\chi}} = 3(2.2)$. Here, $p$ is a proton.
   One can see the  NA64 experiment gives the strongest bound at $m_{\chi} \leq 0.05~GeV $  }
\end{center}
\label{fig01}
%\labelf{fig01}
\vspace{-5mm}
\end{figure}

  \section{Conclusion
  \label{ConclSect}}
  In this paper we have demonstrated that there is a strong complementarity of  searches for LDM with accelerator-based and direct detection experiments.
  Using the latest  NA64 results we compare the discovery perspectives of LDM with these two approaches. 
  As one can see from Fig.1,  for $A'$  masses  $ m_{A`} \leq 300~MeV $  the projections for  NA64 look more
  optimistic in comparison with direct dark matter detection experiments. Moreover,  for
  the model with the vector mediator  and Majorana type  DM,  the cross section
  $\sigma(\chi e(p) \rightarrow \chi e(p) ) $ is suppressed by the factor  $k_M = \frac{2\mu_{\chi e}^2}{m^2_{\chi}}v^2_{\chi}
  = O(10^{-6}) $ that makes the observation  of  LDM scattering in direct experiments extremely challenging or even
  hopeless, and the accelerator searches,  like NA64,  more attractive due to absence of such suppression factor.
  Also for the model with pseudo-Dirac LDM the pseudoelastic reaction
  $\chi_1 e(p) \rightarrow \chi_2 e(p) $ is impossible for $\Delta =|m_{\chi_2} - m_{\chi_1}|  \geq O(10^{-12})$
  due to kinematics considerations, while NA64 and other accelerator experiments like BaBar or BELLE don't have
  such suppression factors.

  \section{Acknowledgements}
  This work is  supported
by the  Russian Science Foundation  RSF grant 21-12-00379. We are indebted to our colleagues from NA64 and INR RAS theoretical depertment for discussion and comments. 
%  \newpage
%  \newpage

\end{document}